\documentstyle[12pt]{article}
\def\headline={\ifnum\pageno=1\firstheadline\else
\ifodd\pageno\rightheadline \else\leftheadline\fi\fi}
\def\firstheadline{\hfil}
\def\rightheadline{\hfil}
\def\leftheadline{\hfil}
\def\footline={\ifnum\pageno=1\firstfootline\else\otherfootline\fi}
\def\firstfootline{\rm\hss\folio\hss}
\def\otherfootline{\hfil}

\font\twelvebf=cmbx10 scaled\magstep 1
\font\twelverm=cmr10 scaled\magstep 1
 1

\font\tenrm=cmr10
\font\tenit=cmti10

\input{epsf}

\begin{document}

\centerline{\twelvebf LEE-$\!$YANG ZEROS FOR SUBSTITUTIONAL SYSTEMS}
\vglue 0.8cm
\centerline{\twelverm HARALD SIMON, MICHAEL BAAKE}
\baselineskip=13pt
\centerline{\tenit Institut f\"ur Theoretische Physik, 
Universit\"at T\"ubingen,}      
\baselineskip=12pt
\centerline{\tenit Auf der Morgenstelle 14, 72076 T\"ubingen, Germany}
\vglue 0.3cm
\centerline{\tenrm and}
\vglue 0.3cm
\centerline{\twelverm UWE GRIMM}
\centerline{\tenit Instituut voor Theoretische Fysica, 
Universiteit van Amsterdam,}
\baselineskip=12pt
\centerline{\tenit Valckenierstraat 65, 1018 XE Amsterdam, The Netherlands}
\vglue 0.8cm
\centerline{\tenrm ABSTRACT}
\baselineskip=13pt
\centerline{\parbox[t]{33.2pc}{\tenrm\baselineskip=12pt
Qualitative and quantitative information about critical 
phenomena is provided 
by the distribution of zeros of the partition function in the complex plane. 
We apply this idea to Ising models on non-periodic systems based on 
substitution. In 1D we consider the Thue-Morse chain and
show that the magnetic field zeros are filling a Cantor subset of the 
unit circle, the gaps being related to a general gap labeling theorem.
In 2D we study the temperature zeros of the Ising model on the 
Ammann-Beenker tiling. The use of corner transfer matrices allows an efficient 
calculation of the partition function for rather large patches which results 
in a reasonable estimate of the critical temperature.}}
\vspace{0.2truein}
\baselineskip=14pt
    
\subsubsection*{1. Introduction}

\mbox{\indent}In 1952, Lee and Yang presented a new approach to questions like 
the existence and location of critical points$^{\ref{Lee}}$. They proposed to
treat the field or fugacity as complex variables and to investigate 
the zeros of the partition function in the complex plane.
Later, also the zeros in the complex temperature plane were studied, they 
also yield information about phase boundaries as well as critical exponents. 
For regular isotropic lattices, these zeros typically lie on simple curves
(though they can fill 2D regions in anisotropic cases).
But for hierarchical graphs, they generally form fractal 
structures$^{\ref{Derrida}}$. Non-periodic graphs with inflation symmetry may
be regarded as a link between the extensively studied regular and 
hierarchical models. So one expects them to combine aspects of both classes. 
This is why we study the Ising model on certain aperiodic graphs
based on substitution rules.

For the classical 1D Ising model on the Thue-Morse chain we will show the 
appearance of fractal structures in the patterns of {\em magnetic field 
zeros} -- a consequence of non-commuting transfer matrices for this
special aperiodic order.
In 2D, however, these zeros did not show any interesting structures$^{\ref{BGP}}$.
Therefore, we present the {\em temperature zeros} of an Ising model on a quasiperiodic 
graph, namely the so-called Ammann-Beenker tiling. 
Its symmetry and the technique of corner transfer matrices 
allow an efficient numerical treatment of quite large patches. 

\subsubsection*{2. Ising Model on the Thue-Morse Chain}

\mbox{\indent}Let us start with the discussion of a 1D chain of $N$ Ising spins 
$\sigma_j\in\{\pm1\}$ with 
periodic boundary conditions ($\sigma_{N+1}=\sigma_1$). 
The energy of a configuration \mbox{\boldmath $\sigma$} reads:
\begin{equation}
E\left(\mbox{\boldmath $\sigma$}\right)\; =\;
-\sum^N_{j=1} 
  \left( J_{j,j+1} \sigma_j \sigma_{j+1}\; +\; H_j \sigma_j \right) \; .
\end{equation}
Here, we consider a system with uniform magnetic field $H_{j}=H$
where the couplings $J_{j,j+1}$ take only two 
different values ${J_a,J_b}$ according to the two letters of the
Thue-Morse chain. The latter is obtained through the substitution rule
\begin{equation}
   S:\begin{array}{c}a\rightarrow ab\\b\rightarrow ba\end{array}
\end{equation}
where we consider the successive periodic systems obtained from
the words $a$, $ab$, $abba$, $abbabaab$ and so on by cyclic closure.
The partition function may be written as the trace of 
$2 \!\times \!2$ transfer matrices$^{\ref{Baxter}}$.
Introducing the notation $z_{a,b}=\exp{(2\beta J_{a,b})}$ and $w=\exp{(2\beta H)}$,
where $\beta$ is the inverse temperature,
the two elementary transfer matrices $T_a$ and $T_b$ read:
\begin{equation}
T_{a,b}= (w z_{a,b})^{-1/2} \left( \begin{array}{cc} 
w z_{a,b} & w^{\scriptscriptstyle 1/2} \\ 
w^{\scriptscriptstyle 1/2} & z_{a,b} \end{array} \right) \; .
\end{equation}
The recursion relation for the chain is the same as that for the 
transfer matrices, and so the essential part 
of the partition function is evidently a
polynomial$^{\ref{BGJ},\ref{BGP}}$ in the three variables $z_a,z_b$ and $w$.

In what follows, we restrict ourselves to the ferromagnetic 
regime (i.e.\ $z_a, z_b \ge 1$), focusing
on the magnetic field zeros for fixed positive temperature. 
Due to the quite general Lee-$\!$Yang theorem$^{\ref{Lieb}}$
the magnetic field zeros are restricted to the unit circle. For a periodic
chain the zeros can be calculated analytically$^{\ref{Pathria}}$, 
where they fill a connected part of the
unit circle densely in the thermodynamic limit. 
The only gap is near the positive real axis, due to the fact that 
there is no phase transition for finite temperature in 1D.  
\begin{figure}[h]
\label{tmzeros}
\centerline{\epsfysize=66mm\epsfbox[90 300 520 500]{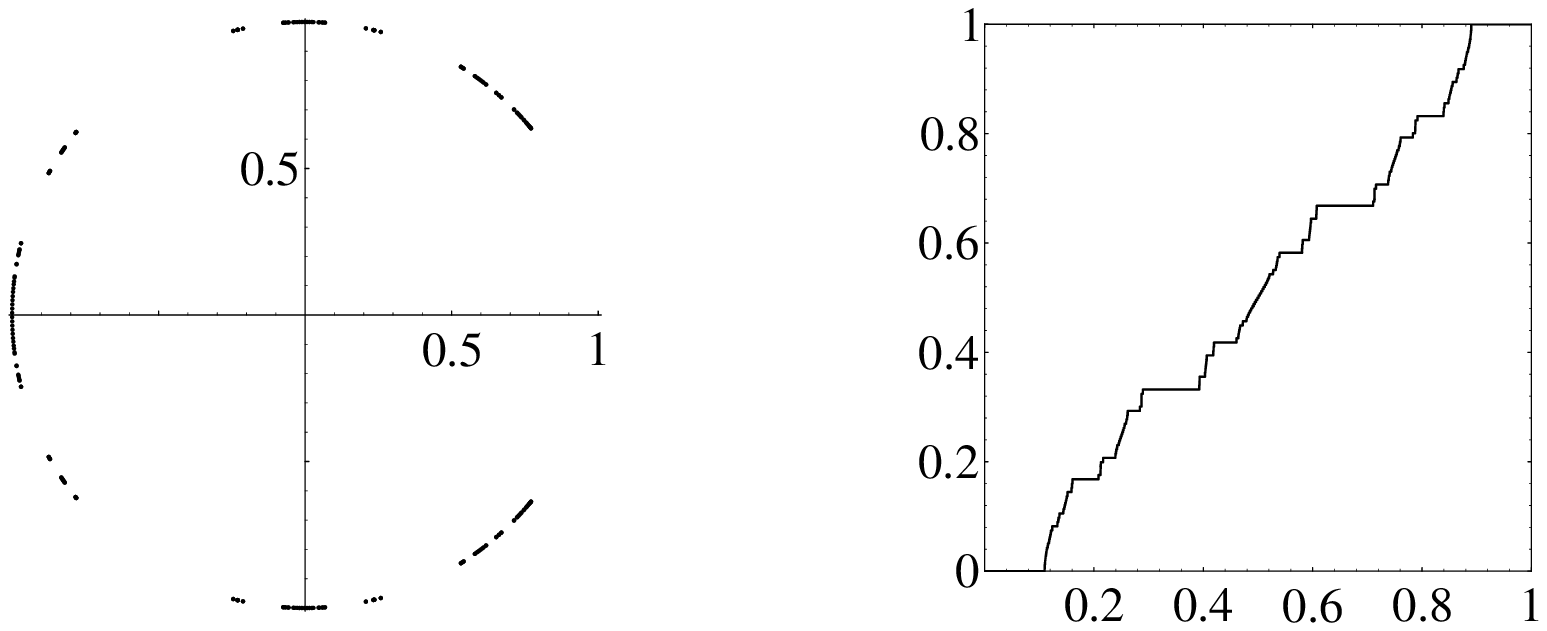}}
\centerline{\tenrm Fig.~1. Magnetic field zeros and their integrated density
 on the unit circle.}
\end{figure}
In Fig.~1 we show the magnetic field zeros of the Thue-Morse chain
in the $w$-plane in the ferromagnetic case $z_a=3/2$, $z_b=100$ for the periodic
approximant of length $2^8=256$. 
As expected, the gap around the real axis
near the point $w=1$ is still present. But there is, in fact, an infinite
hierarchy of gaps, each with the well-known Lee-$\!$Yang edge singularity. 
It is an interesting property that these gaps (through the definition of a discrete
step function along the unit circle)
may be related to the gap labeling in the electronic
spectrum of the Thue-Morse chain, for details see Refs.~\ref{BGJ} and \ref{Belli}.

\subsubsection*{3. Ising Model on the Ammann-Beenker Tiling}

\mbox{\indent}Exact results for 2D quasiperiodic models are rather rare and 
generally restricted to very special cases. Even for systems with an 
inflation symmetry no exact
renormalization is known for electronic systems or Ising models. 
So, we apply a combination of an exact calculation of a 
finite partition function followed by an investigation of the 
complex zeros as an approach to the thermodynamic limit.
\begin{figure}[ht]
\centerline{\epsfbox[10 5 140 150]{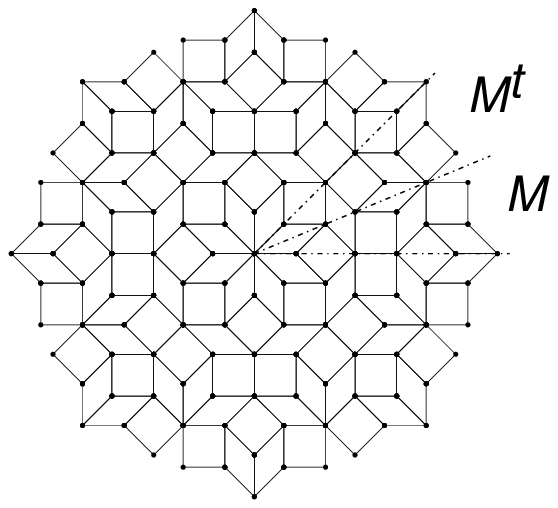}}
\centerline{\tenrm Fig.~2.  Octagonal patch of the Ammann-Beenker tiling.}
\end{figure}

For our 2D quasiperiodic Ising model, we have chosen the 
Ammann-Beenker tiling$^{\ref{Ammann}}$.
It has only one kind of edges, which suggests a simple choice 
of equal couplings along all bonds.
But it is even more important that the octagonal symmetry allows 
the application of corner transfer matrices$^{\ref{Baxter}}$.
It may be built repeating the indicated small sector $16$ times. 
The corresponding (rectangular) corner transfer matrix
$\mbox{\boldmath $M$}$ is easy to calculate. Consequently
the partition function $Z\left(w,z\right)$ is simply given by:
\begin{equation}
Z\left(w,z\right) = tr\left(  
(\mbox{\boldmath $M$}^t \mbox{\boldmath $M$})^8 \right) \; .
\end{equation}
This simple structure allows the exact calculation of the partition function 
for large patches using algebraic manipulation packages.
So our calculations were limited by the degree of the resulting 
polynomial partition function, as the numerical calculation of 
polynomial roots quickly becomes really involved.

In contrast to 1D, the magnetic field ($w$) zeros do not seem to contain 
any relevant new information:
their angular distribution looks astonishingly regular in
comparison to the 1D case, and no gap structure is visible.
So, we concentrate on the temperature ($z$) zeros here and
\begin{figure}[h]
\label{abzeros}
\centerline{\epsfbox[220 320 400 480]{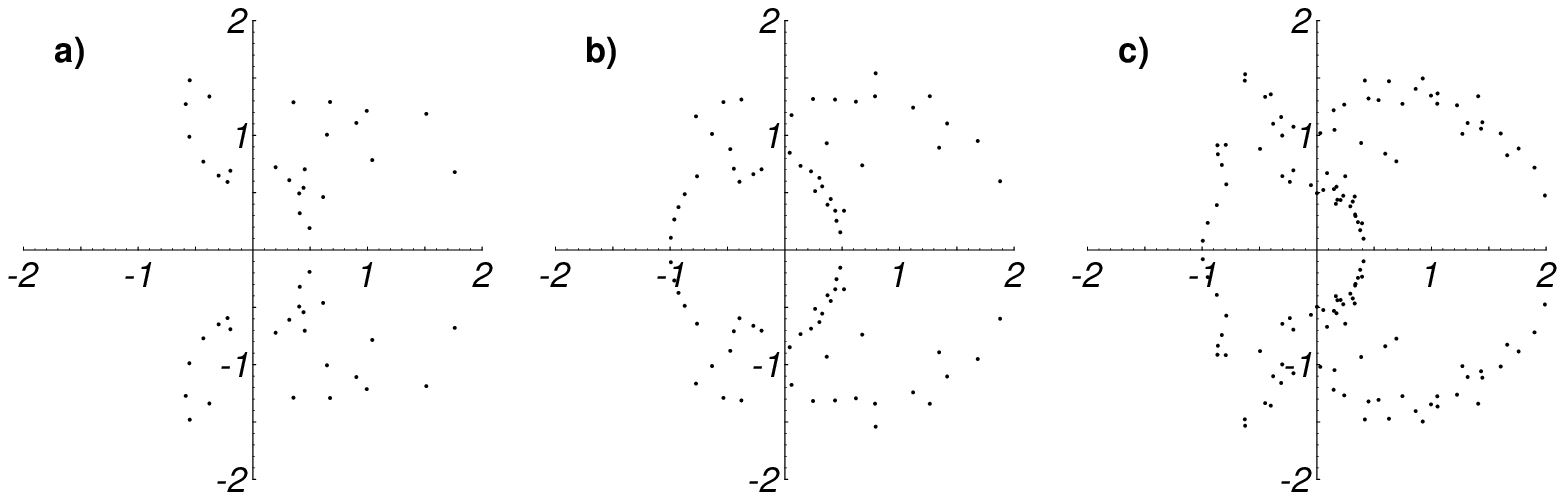}}
\centerline{\tenrm Fig.~3. Partition function zeros in the complex 
temperature plane.}
\end{figure}
restrict ourselves to the case of zero magnetic field. 
In Fig.~3, the temperature zeros are shown for a growing sequence of 
patches with fixed boundary conditions, with Fig.~3c) corresponding to 
the patch shown in Fig.~2.
In principle the zeros do not appear to lie on simple curves. But those
near the real axis directly contain information about the critical point 
and indirectly even about critical exponents (for which one would need 
to know the dependence of the zeros on the magnetic field).
Fig.~3 shows alignments of zeros ($Re(z)>1$) converging towards 
two points of the real axis. The numerical values of the zeros
closest to the critical ferromagnetic as well as antiferromagnetic couplings are given 
in Table~1. In the ferromagnetic case they  are in  very good agreement with
the specific heat and center spin magnetization also obtained 
by numerical calculations. In comparison to the square lattice
(where the critical coupling is 
$z_{c}=1+\sqrt{\mbox{\small 2}}$), 
the local coordination looks a bit higher (though its average is strictly 4), 
and the critical coupling shows this by a slight decrease in agreement with 
other results$^{\ref{Ledue}}$. 
As our graph is bipartite, the critical antiferromagnetic coupling is just
the reciprocal of the ferromagnetic one. Due to the fixed boundary conditions,
which are not appropriate for the antiferromagnetic case, we expect 
our numerical values to show large finite-size effects there. 

\begin{table}[h]
\centerline{\tenrm Table 1. Zeros of the partition function closest to the real axes.}
\vspace{2mm}
\centerline{\begin{tabular}{|c|c|c|c|}
\hline
&a) & b) & c) \\
\hline
Ferromagnetic: &
$ 1.7608 \pm 0.6795 i $ & $ 1.8772 \pm 0.5993 i $ & $ 1.9895 \pm 0.4752 i $\\
\hline
Antiferromagnetic: &
$ 0.4944 \pm 0.1906 i $ & $ 0.4834 \pm 0.1543 i $ & $ 0.4075 \pm 0.0988 i $\\
\hline
\end{tabular}}
\end{table}
 
\vspace{-3mm}

\subsubsection*{4. Conclusion}

\mbox{\indent}Our 1D calculations showed the fractal structure of the magnetic field zeros for 
1D non-periodic Ising models, while this structure seems to be absent in 2D. 
This and the relation to a gap labeling is quite similar to the electronic and 
vibrational spectra. For 2D Ising like systems the investigation of partition 
function zeros yields valuable information about the critical point of 
models on quasiperiodic graphs. 
Their distribution is rather more complicated than for regular periodic 
graphs but the location of the ferromagnetic phase transition clearly 
shows up where the zeros "pinch" the real axis. 

\subsubsection*{5. References}

{\small
\begin{enumerate}
\baselineskip=12pt
\parskip=0.9pt

\newcommand{\bibi}{\item}

\bibi \label{Ammann}
R.~Ammann, B.~Gr\"{u}nbaum and G.~C.~Shephard,
{\em Aperiodic tiles}, Discrete Comput.\ Geom.\ {\bf 8} (1992) 1-25.

\bibi \label{BGJ}
M.~Baake, U.~Grimm and D.~Joseph,
``Trace Maps, Invariants, and Some of their Applications'',
{\em Int.\ J.\ Mod.\ Phys.} {\bf B7} (1993) 1527--50;
and: ``Practical Gap Labeling'', in preparation.

\bibi \label{BGP}
M.~Baake, U.~Grimm and C.~Pisani,
``Partition Function Zeros for Aperiodic Systems'',
{\em J.\ Stat.\ Phys.} {\bf 78} (1995) 285--97.

\bibi \label{Baxter}
R.~J.~Baxter,
{\em Exactly Solved Models in Statistical Mechanics},
Academic Press, London (1982).

\bibi \label{Belli}
J.\ Bellissard,
``Spectral Properties of Schr\"odinger's Operator
with a Thue-Morse Potential'',
in: {\em Number Theory and Physics}, eds. J.-M.\ Luck,
P.\ Moussa and M.\ Waldschmidt, Springer Proceedings in Physics,
vol.\ 47, Springer, Berlin (1990), p.~140--50.

\bibi \label{Derrida}
B.~Derrida, L.~De Seze and C.~Itzykson,
``Fractal Structure of Zeros in Hierarchical Models'',
{\em J.\ Stat.\ Phys.} {\bf 33} (1983) 559--69.

\bibi \label{Ledue}
D.~Ledue, D.~P.~Landau and J.~Teillet,
``Static critical behaviour of the ferromagnetic Ising model
on the quasiperiodic octagonal tiling'',
{\em Phys.\ Rev.} {\bf B51} (1995) 12523--30.

\bibi \label{Lieb}
E.~H.~Lieb and A.~D.~Sokal,
``A General Lee-$\!$Yang Theorem for One-Component and Multicomponent Ferromagnets'',
{\em Commun. Math. Phys.} {\bf 80} (1981) 153-79.

\bibi \label{Lee}
T.~D.~Lee and C.~N.~Yang,
``Statistical Theory of Equations of State and Phase Transitions.\
  II. Lattice Gas and Ising Model'',
{\em Phys.\ Rev.} {\bf 87} (1952) 410--9.

\bibi \label{Pathria}
R.~K.~Pathria,
{\em Statistical Mechanics}, Pergamon, Oxford (1972).

\end{enumerate}
 }
\end{document}